\documentclass{emulateapj}

\newcommand{\mps}{m s$^{-1}$}
\newcommand{\kmps}{km s$^{-1}$}

\shorttitle{RADIAL VELOCITY VARIABILITY IN HUBBLE I 4}
\shortauthors{MAHMUD ET AL.}

\begin{document}

\slugcomment{Draft version; accepted for publication in the Astrophysical Journal, May 2011}

\title{STARSPOT-INDUCED OPTICAL AND INFRARED RADIAL VELOCITY VARIABILITY IN T TAURI STAR HUBBLE I 4}

\author{Naved I. Mahmud\altaffilmark{1,2,3}, Christopher J. Crockett\altaffilmark{2,3,4,5}, Christopher M. Johns-Krull\altaffilmark{1,2,3}, L. Prato\altaffilmark{2,3,4},\\
Patrick M. Hartigan\altaffilmark{1}, Daniel T. Jaffe\altaffilmark{6}, and Charles A. Beichman\altaffilmark{7,8}}

\altaffiltext{1}{Rice University, 6100 Main Street, MS 108, Houston, TX 77005; naved@rice.edu, cmj@rice.edu, hartigan@rice.edu.}
\altaffiltext{2}{Visiting Astronomer at the McDonald Observatory of the University of Texas at Austin.}
\altaffiltext{3}{Visiting Astronomer at the Infrared Telescope Facility, which is operated by the University of Hawaii under Cooperative Agreement no. NNX-08AE38A with the National Aeronautics and Space Administration, Science Mission Directorate, Planetary Astronomy Program.}
\altaffiltext{4}{Lowell Observatory, 1400 West Mars Hill Road, Flagstaff, AZ 86001; crockett@lowell.edu, lprato@lowell.edu.}
\altaffiltext{5}{University of California Los Angeles, 430 Portola Plaza, Box 951547, Los Angeles, CA 90095-1547.}
\altaffiltext{6}{University of Texas, Robert Lee Moore Hall, Austin, TX 78712; dtj@astro.as.utexas.edu.}
\altaffiltext{7}{Jet Propulsion Laboratory, California Institute of Technology, 4800 Oak Grove Drive,
Pasadena, CA 91109; chas@pop.jpl.nasa.gov.}
\altaffiltext{8}{NASA Exoplanet Science Institute (NExScI), California Institute of Technology, 770 South Wilson Avenue, Pasadena, CA 91125.}

\begin{abstract}
We report optical ($\sim$6150 \AA) and K-band (2.3 $\mu$m) radial velocities obtained over two years for the pre-main sequence weak-lined T Tauri star Hubble I 4. We detect periodic and near-sinusoidal radial velocity variations at both wavelengths, with a semi-amplitude of $1395\pm94$ \mps~in the optical and $365\pm80$ \mps~in the infrared. The lower velocity amplitude at the longer wavelength, combined with bisector analysis and spot modeling, indicates that there are large, cool spots on the stellar surface that are causing the radial velocity modulation. The radial velocities maintain phase coherence over hundreds of days suggesting that the starspots are long-lived. This is one of the first active stars where the spot-induced velocity modulation has been resolved in the infrared. 
\end{abstract}


\section{INTRODUCTION}

One of the most exciting and important contributions made by astronomers in the recent past has been the discovery of planets outside of our Solar System. In the past two decades, over 500 such extrasolar planets have been discovered (Extrasolar Planets Encyclopedia; exoplanet.eu). The Kepler Mission is likely to increase this number by an order of magnitude or more. Of the extrasolar planets currently confirmed, the vast majority have been detected using the radial velocity (RV) method, which uses Doppler spectroscopy to measure the tiny wobble of the host star caused by an orbiting planet. The advent of Kepler notwithstanding, the RV method is likely to remain the most important technique for confirming extrasolar planets in the coming decade. The RV method will also play an important role in the search for nearby habitable planets, one of the primary objectives of the astronomical community as identified in the 2010 Decadal Survey (New Worlds, New Horizons in Astronomy and Astrophysics; National Research Council). Thus it is an important goal to understand and characterize the various sources of uncertainties that the RV method is susceptible to.

There are a number of astrophysical sources of RV noise (e.g. acoustic p-modes, meridional flows, starspots). Spots on the surface of a rotating star add RV noise by causing temporal changes in the profiles of the photospheric absorption lines used to measure Doppler shifts (e.g. Vogt \& Penrod 1983). Spots are especially problematic because not only do they add RV jitter, but spots that persist for timescales much greater than the stellar rotation period can also cause the false-positive detection of a companion. This false detection originates because a spot that is visible at all times on the surface of an inclined star gives rise to apparent RV modulation that closely mimics the periodic RV signal of a companion (Queloz et al. 2001; Huerta et al. 2008).

Several authors have determined empirical relationships between stellar parameters and the level of spot-induced RV jitter. Saar et al. (1998) found a relationship between stellar rotation velocity ($v$sin$i$) and RV noise. Wright (2005) measured a relationship between stellar magnetic activity (the $S$ index, determined from chromospheric Ca II H and K lines) and RV jitter based on $\sim$450 stars, with more active stars showing greater RV noise. These relationships can be used to predict the level of RV noise in a given star; however, they are not sufficient for determining the origin of all observed RV variability.

Various indicators have been used to distinguish if the observed RV modulation of a star is spot-induced or companion-induced. However, no indicator has been demonstrated to be consistently reliable. Photometric monitoring can be used to measure the rotation period of a spotted star. If the rotation period is the same as the RV orbital period, then it is likely that the RV variability is spot-induced; however, similar periods could also be the result of a close-in, tidally-locked planet (Marcy et al. 1997). A similar approach of looking for identical rotation and orbital periods can be used with stellar activity indicators. Queloz et al. (2001) ruled out a planet around HD 166435 in part using the rotational modulation of the magnetic activity indicator $S$. However, RVs and magnetic activity are not always correlated. Wright et al. (2008) found no periodic RV variations in four stars that show significant stellar activity cycles coherent over several years; however, another of their targets (HD 154345) does show a very strong correlation between RV and $S$.

Bisector analysis has also been used to determine the origin of RV variability. A spot will distort the profile of stellar absorption lines, changing the asymmetry of each line. As the star rotates, the spot crosses from one limb of the star to the other, producing a continuous sequence of distortions across the line profile. Spectra of the star at two distinct rotational phases provide snapshots of the distortion, quantified by the line bisector span: the difference in bisector value at two different locations in the line profile. The spot variations result in an apparent RV shift. The size of this shift is expected to be correlated with the size of the line bisector span. Bisector analysis was used to strengthen the case for the discovery of the first exoplanet around a Sun-like star, 51 Pegasi (Mayor \& Queloz 1995; Hatzes et al. 1997), and has subsequently become a critical tool for identifying misleading RV signals caused by spots. However, similar to stellar activity analysis, bisector analysis does not always work for determining the origin of RV variability (e.g. Hu{\'e}lamo et al. 2008). Prato et al. (2008) showed examples of young stars that do not display significant correlations between their bisector spans and RVs, yet their RV variations likely originate in spots. Desort et al. (2007) showed that when $v$sin$i$ is lower than the resolution of the spectrograph, RV and bisector span variations do not correlate, mimicking the expected behavior of planetary companions (e.g. Setiawan et al. 2008). Thus, even though it is often useful to look for correlations between RV variability and photometric, stellar activity, or bisector span variability, there is no indicator that can consistently be used to uncover false-positive detections of companions.

Another way to mitigate the effect of spots is to observe at well-separated multiple wavelengths. The contrast between a photosphere and a cooler starspot decreases at redder wavelengths because of the flux-temperature scaling in the Rayleigh-Jeans limit of blackbody radiation (e.g. Vrba et al. 1986). As a result, the amplitude of any spot-induced RV variability will be smaller at longer wavelengths, whereas the reflex motion caused by a true companion will be the same at all wavelengths. Thus in recent years there has been an increasing interest in near-infrared (NIR) RV searches.

Surveys in the NIR are especially important for finding planets around young stars, crucial targets for putting observational constraints on the various planet formation theories (e.g. core accretion versus gravitational instabilities). Low-mass young stars (T Tauri stars) have very strong magnetic fields (e.g. Johns-Krull 2007) that can generate multiple large cool spots. Such spotted stars are expected to have very high levels of spot-induced RV jitter (Saar \& Donahue 1997). Several surveys have focused on young stars but most have not identified planets because of high levels of RV noise in the optical and small sample sizes. Paulson et al. (2004) searched for planetary companions to stars in the Hyades ($\sim$790 Myr) but found none. Paulson \& Yelda (2006) studied 61 nearby stars with ages $12-300$ Myr and found no planets with masses $>$$1-2$ $M_{Jup}$ at the 3$\sigma$ level. Setiawan et al. (2007) detected a $>$6.1 $M_{Jup}$ planet around a 100 Myr old star with a 852 d period. Setiawan et al. (2008) also reported a 10 $M_{Jup}$ planet around a 10 Myr old T Tauri star (TW Hya); however, based in part on discrepant RV measurements at optical and NIR wavelengths, Hu{\'e}lamo et al. (2008) attributed the RV signal from this star to spots. Similarly, Prato et al. (2008) used optical and NIR RVs to rule out substellar companions around the young stars DN Tau and V836 Tau. Mart\'{i}n et al. (2006), based on their optical and infrared RV observations of a brown dwarf, also stressed the need for multi-wavelength observations to verify the origin of any periodic RV variability. Thus infrared surveys will be essential for finding the youngest planets and refining the theories of planet formation.

Apart from the promise of lower spot-induced noise, infrared surveys are also being initiated to look for planets around M dwarfs; these stars have SEDs that peak in the infrared. Most surveys have focused on optically bright FGK dwarfs owing to the greater availability and technological maturity of high-precision spectrographs that operate in the optical. However, M dwarfs are the most numerous stars in the Milky Way (Covey et al. 2008) and will have greater reflex motions from a given planet than higher-mass stars. Their habitable zones will also be closer in because of their lower luminosities (Kasting et al. 1993) and planets in this region will produce correspondingly higher RV
amplitudes than in the habitable zone of a higher-mass star. Thus various groups have begun infrared RV surveys that focus on low-mass stars (Bean et al. 2010; Blake et al. 2010).

However, infrared RV observations are not immune to the effect of spots. In particular, the low-mass stars that are the focus of current infrared surveys are more likely to be magnetically active (West et al. 2004). However, it is unknown what the intrinsic infrared RV variability is for active stars (T Tauri stars or M dwarfs). This will depend on the topology, size, and temperature of the resulting spots. The exact relationship between mass, age, and activity is also currently unclear. Thus it is crucial to study active stars of different types at multiple wavelengths to accurately characterize the effect of magnetic activity on RV measurements. Such studies will also allow us to learn more about spots and dynamos. In this paper, we report the optical and NIR RV variability observed for the T Tauri star \object{Hubble I 4}. Hubble I 4 is a weak-lined T Tauri star (WTTS) of spectral type K7; it is also a Class III object, i.e. it has very little infrared excess in its SED, indicating a fully-dissipated circumstellar disk (Furlan et al. 2006; White \& Ghez 2001). Johns-Krull et al. (2004) measured $T_{eff}=4158\pm56$ K, $v$sin$i=14.6\pm1.7$ \kmps, and a mean magnetic field of $2.51\pm0.18$ kG for Hubble I 4. Kraus et al. (2011) recently identified Hubble I 4 as a binary using high-resolution imaging, with a separation of 4.1 AU. An orbital solution with 4 epochs indicates an inclination angle of $\sim$$20^{\circ}$ and a period of $\sim$9 years (Kraus 2011, priv. comm.). Velocity variations expected as a result of this companion are on a much larger timescale than investigated in this study. The amplitude of such variations will also be quite small owing to the long orbital period and the low orbital inclination.

This paper is organized as follows. In \S2 we describe our observations and data reduction. In \S3 we discuss our RV measurements, period determination, and bisector analysis. We discuss our results in \S4 and provide a conclusion in \S5. 

\section{OBSERVATIONS AND DATA REDUCTION}

Hubble I 4 is part of our ongoing survey of young stars in search of giant planet and brown dwarf companions (Huerta et al. 2008; Prato et al. 2008). Our total sample consists of 140 G, K, and M type stars in the nearby ($\sim$140 pc), young (1$-$few Myr) Taurus-Auriga association, and in the Pleiades open cluster ($\sim$100 Myr). We selected our T Tauri target sample from the Herbig and Bell catalogue (Herbig \& Bell 1988) based on the following criteria: (1) $v$sin$i<20$ \kmps~to maximize the chances of observing strong, narrow absorption lines, (2) $V<15$, (3) no binaries with separations $<$0\arcsec.05 to avoid confusion in the velocity signature, and (4) G, K, or M spectral type to maximize the number of spectral lines in the red and thus the RV precision. Applying these criteria yielded a sample of $\sim$50 classical (actively accreting) and $\sim$90 weak-lined (weakly or not accreting) T Tauri stars. Among the classical T Tauris in our sample, none are so strongly veiled as to render the absorption lines too weak to measure.

Our observing strategy for this survey consists of obtaining every-night observations of our targets for roughly a week at a time. The week-long observing window is chosen to closely match both the typical rotation periods of T Tauri stars and the companion orbital periods we are capable of detecting. Typically, we also get a second week of observations for each target separated by about two months; thus, we are also sensitive to longer periods.

\subsection{Optical Spectroscopy}

We took optical spectra of Hubble I 4 at McDonald Observatory using the 2.7-meter Harlan J. Smith Telescope and the Robert G. Tull Coud\'{e} Spectrograph (Tull et al. 1995). We obtained 26 spectra of Hubble I 4 between 2008 and 2010 spanning 447 days (Table 1). A 1\arcsec.2 slit yielded a spectral resolving power of $R\sim60,000$. Integration times were typically 1800 s (depending on conditions) and average seeing was $\sim$2\arcsec. We took ThAr lamp exposures before and after each spectrum for wavelength calibration; typical RMS values for the dispersion solution precision were $\sim$4 \mps. We observed RV standards on every night; their overall RMS scatter is $\sim$140 \mps~(details in \S3.1). 

\begin{deluxetable}{ccc}
\tablewidth{0pt}
\tablecaption{Hubble I 4 Optical Radial Velocities}
\tablehead{
\colhead{JD} & \colhead{Radial Velocity (\kmps)} & \colhead{$\sigma$ (\kmps)}
}
\startdata
2454788.674 & 2.200 & 0.181 \\
2454789.675 & 2.940 & 0.171 \\
2454790.672 & 0.530 & 0.167 \\
2454792.668 & 2.304 & 0.192 \\
2454793.678 & 0.799 & 0.205 \\
2454794.665 & 2.703 & 0.186 \\
2454795.675 & 1.906 & 0.207 \\
2454842.619 & 2.039 & 0.163 \\
2454843.608 & 1.373 & 0.156 \\
2454844.593 & 0.063 & 0.150 \\
2454845.585 & 2.701 & 0.157 \\
2454846.587 & 0.926 & 0.169 \\
2454847.588 & 0.633 & 0.165 \\
2454848.592 & 2.455 & 0.166 \\
2454849.579 & 0.422 & 0.183 \\
2455159.777 & 0.595 & 0.158 \\
2455160.761 & 2.592 & 0.157 \\
2455161.730 & 0.000 & 0.140 \\
2455162.764 & 1.584 & 0.163 \\
2455163.848 & 2.675 & 0.254 \\
2455164.902 &-0.144 & 0.224 \\
2455229.581 &-0.055 & 0.229 \\
2455232.581 & 0.236 & 0.168 \\
2455233.564 & 3.087 & 0.175 \\
2455234.564 & 0.850 & 0.196 \\
2455235.564 & 1.055 & 0.189 \\
\enddata
\end{deluxetable}

All optical data reduction was performed using IDL routines written for reducing echelle spectra. These routines form the basis of the REDUCE IDL echelle reduction package (Piskunov \& Valenti 2002). The raw spectra were bias-subtracted using the overscan region and flat-fielded using the spectrum of a continuum internal lamp. Optimal extraction to remove cosmic rays and improve signal was used for all the spectra. Wavelength calibration was done by averaging the wavelength solution from ThAr lamp exposures taken before and after each stellar exposure.\\\\

\subsection{Infrared Spectroscopy}

We obtained NIR (K-band) spectra of Hubble I 4 at the 3-meter NASA Infrared Telescope Facility (IRTF) using the high-resolution, Cassegrain-mounted, echelle spectrograph, CSHELL (Tokunaga et al. 1990; Greene et al. 1993). We observed 18 spectra between 2008 and 2010 spanning 462 days (Table 2). We used the Circular Variable Filter (CVF) to isolate a 50 \AA~segment of spectrum centered at 2.298 $\mu$m onto the $256\times256$ InSb detector array. A 0\arcsec.5 slit yielded a FWHM of typically 2.6 pixels ($\sim$0.5 \AA) corresponding to a spectral resolving power of $R\sim46,000$.

\begin{deluxetable}{ccc}
\tablewidth{0pt}
\tablecaption{Hubble I 4 Infrared Radial Velocities}
\tablehead{
\colhead{JD} & \colhead{Radial Velocity (\kmps)} & \colhead{$\sigma$ (\kmps)}
}
\startdata
2454780.851 & 17.418 & 0.150 \\
2454782.819 & 16.706 & 0.211 \\
2454783.794 & 17.159 & 0.201 \\
2454785.838 & 16.815 & 0.219 \\
2454786.847 & 17.334 & 0.121 \\
2455151.094 & 16.063 & 0.247 \\
2455153.119 & 16.877 & 0.191 \\
2455156.054 & 16.693 & 0.104 \\
2455158.052 & 16.977 & 0.095 \\
2455160.052 & 16.078 & 0.102 \\
2455235.830 & 16.426 & 0.139 \\
2455236.798 & 17.178 & 0.176 \\
2455237.808 & 16.747 & 0.273 \\
2455238.826 & 16.508 & 0.095 \\
2455239.844 & 17.124 & 0.171 \\
2455240.827 & 16.734 & 0.154 \\
2455241.802 & 16.981 & 0.165 \\
2455242.800 & 17.081 & 0.098 \\
\enddata
\end{deluxetable}

We estimated the seeing by fitting each column (the cross-dispersion direction) of our bright standard star spectra to a Gaussian function. We calculated the median FWHM across all columns for each exposure, and then determined the median for all standard star spectra across all nights. Using this approach, we estimated that the seeing was typically $\sim$1\arcsec.18. However, because of peculiarities in the image quality and telescope jitter over the course of the exposures, this is likely to be an overestimate.

Flat fields were imaged each night using a continuum lamp to illuminate the entire slit. At the start of each night, we imaged six Ar-Kr-Xe emission lines by changing the CVF while maintaining the grating position to determine the wavelength reference. We found the dispersion solution by fitting a third order polynomial to the locations of these emission lines. All of our target data were obtained using 10\arcsec~nodded pairs to enable subtraction of sky emission, dark current, and detector bias. In addition to Hubble I 4 and several other T Tauri young-planet candidates, we also observed the RV standard GJ 281, known to be stable at a few \mps~(Endl et al. 2003). Total exposure times were $\sim$1 hour for our T Tauri targets (3 nodded pairs; 600 s per nod) and $\sim$20 minutes for the RV standards resulting in a typical signal-to-noise ratio (S/N) of $\sim$$70-150$. Details of the infrared data reduction are given in Crockett et al. (2011).

\begin{figure}
%\plotone{f1.eps}
\includegraphics[scale = 0.5, trim = 0 60mm 0 0, clip]{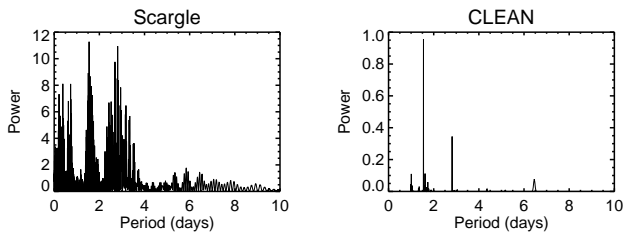}
\caption{Power spectra for Hubble I 4 optical RVs using Scargle (\emph{left}) and CLEAN (\emph{right}). The two strongest peaks are at 1.55 d and 2.81 d.}
\end{figure}

\begin{figure}
%\plotone{f2.eps}
\includegraphics[scale = 0.5, trim = 0 60mm 0 0, clip]{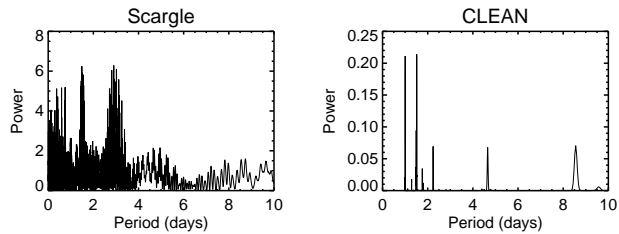}
\caption{Power spectra for Hubble I 4 infrared RVs using Scargle (\emph{left}) and CLEAN (\emph{right}).}
\end{figure}

\section{ANALYSIS}

\subsection{Optical Radial Velocity Measurements}

Following Huerta et al. (2008), we determined optical RVs using a cross-correlation analysis of 9 echelle orders, each covering $\sim$100 \AA. These orders spanned the wavelength range 5600 \AA~to 6700 \AA. The orders were chosen for their high signal-to-noise ratio, lack of stellar emission lines, and lack of strong telluric absorption lines. We used the mean of the RVs derived from the multiple echelle orders as our final RV, while the standard deviation of the mean provided the internal uncertainty. We used the Hubble I 4 observation with the strongest S/N (JD 2455161.730) as the template for the cross-correlation analysis. Using the target itself as the cross-correlation template prevents any spectral-type mismatch that could lead to additional uncertainty in the measurements. Thus our RVs are relative to one observation epoch. All velocities have been corrected for Solar System barycenter motion and are listed in Table 1.

In order to quantify the precision of our optical RV technique, we observed several RV standard stars (107 Psc, HD 4628, $\tau$ Ceti, HD 65277, HD 80367, HD 88371) taken from Nidever et al. (2002), Butler et al. (1996), and Cumming et al. (1999). These are known to be stable at a few \mps. We observed these standards every night that we observed Hubble I 4, and reduced and analyzed them in the same way. The internal uncertainties (from the order-to-order scatter) in our RV standards are quite small ($<$20 \mps); however, the long-term RV scatter of our standard stars over the six years of our survey at McDonald Observatory is $\sim$140 \mps. We identified this larger value as the uncertainty in the method: our nominal level of precision. The final uncertainty for each observation listed in Table 1 is this value (140 \mps) added in quadrature with the internal uncertainty estimated from the order-to-order scatter in the RV determination from the cross-correlation analysis of Hubble I 4.

\subsection{Infrared Radial Velocity Measurements}

We determined NIR RVs using the telluric absorption features in our spectra as an absolute wavelength reference. The details of this technique are described in Crockett et al. (2011). Applying this technique to our RV standard GJ 281 yielded a standard deviation of 58 \mps. The internal uncertainties for this standard, measured from photon statistics, are $\sim$37 \mps. The external error, which we assumed added in quadrature with our internal errors to give the overall scatter in velocities, is therefore 45 \mps. We adopted this value as the uncertainty in the method. Table 2 presents the K-band RV measurements of Hubble I 4; these RVs are measured with respect to the Earth's atmosphere and have been corrected for Solar System barycenter motion. The uncertainties shown in the third column were determined by adding the internal errors from photon statistics in quadrature with the 45 \mps~external error. The varying internal errors from night to night result from variable S/N because of weather conditions. 

\subsection{Detection of Periodic Signals}

To look for periodicity in our RV measurements, we used an IDL implementation of the Scargle method (Scargle 1982) of power-spectrum calculation as implemented by Horne and Baliunas (1986). This is a suitable method for our RVs as it does not require the data to be evenly spaced in time, which ours are not. We also used the discrete Fourier transform plus CLEAN method of Roberts et al. (1987), which attempts to remove the couplings between physical periods and their aliases by deconvolving the spectral window function from the discrete Fourier transform. We performed these analyses on both the optical RVs (26 points) and the NIR RVs (18 points). The power spectra are shown in Figure 1 (optical RVs) and Figure 2 (NIR RVs). The two strongest peaks in the power spectra are at 1.55 d\\\\ and 2.81 d. These periods are present in both the optical RVs and the NIR RVs. We suspect the 2.81 d period to be an alias of the physical 1.55 d period since it is significantly weakened in the CLEAN analysis of the optical data. However, the CLEAN analysis of the NIR data is inconclusive with two equally strong periods.

To test the significance of the 1.55 d period, we performed a Monte Carlo simulation. We constructed 10,000 sets of normally distributed random RV data with overall scatter similar to what we measured in the optical for Hubble I 4. These were analyzed using the Scargle method with the same temporal cadence as our actual observations. None of these power spectra had a peak as strong or stronger than the 1.55 d peak in the Scargle method power spectrum of Hubble I 4. This indicates a high significance for this period detection, and we estimate its false alarm probability (FAP) to be less than $10^{-4}$. We used this period (1.5459 d) to phase-fold the RV data, using the same phase zero-point (JD 2454780.851) for both the optical and the NIR. These RVs are shown in Figure 3.

\begin{figure}
%\plotone{f3.eps}
\includegraphics[scale = 0.5, trim = 0 60mm 0 0, clip]{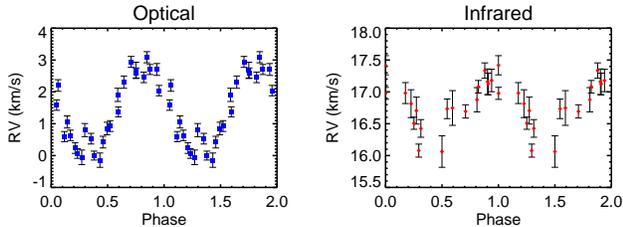}
\caption{Optical (\emph{left}) and infrared (\emph{right}) phase-folded RVs for Hubble I 4, using a period of 1.5459 d. The optical RVs were measured relative to one observation epoch, whereas the infrared RVs were measured with respect to the Earth's atmosphere.}
\end{figure}

\subsection{Bisector Analysis}

We performed bisector analysis on the Hubble I 4 optical data to help determine the origin of its RV variability. For each of the 9 echelle orders used to measure the RV for a given observation, we cross-correlated all absorption lines with the corresponding order in the template spectrum (JD 2455161.730) and measured the cross-correlation function (CCF). We used each CCF (one for each echelle order) to measure the bisector span (the inverse of the mean slope of the bisector) and calculated a mean bisector span for each observation. Many spotted stars show a strong correlation between bisector span and RV (Huerta et al. 2008; Prato et al. 2008). Hubble I 4 shows a similar correlation (Figure 4). The linear correlation coefficient between the bisector spans and the RVs for Hubble I 4 is 0.80 with a FAP of $\sim$$10^{-6}$. This correlation strongly suggests that the RV variability of Hubble I 4 is spot-induced.

\begin{figure}
\plotone{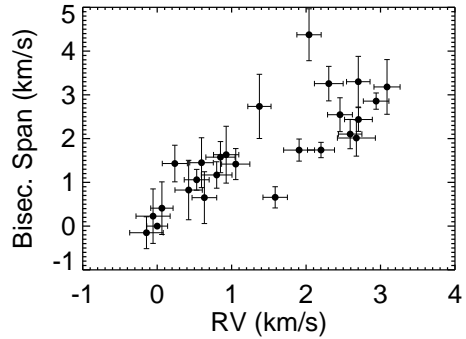}
\caption{Bisector analysis for Hubble I 4. The RVs and bisector spans are correlated with a linear correlation coefficient of 0.80 and a FAP of $\sim$$10^{-6}$.}
\end{figure}

\section{DISCUSSION}

Our analysis provides multiple pieces of evidence that the periodic RV variability seen in Hubble I 4 is the result of spots and not reflex motion caused by a companion. The SuperWASP photometric survey has observed Hubble I 4 over 100 times (spanning several months in 2004), and reported a rotation period of 1.5483 d (Norton et al. 2007). This is almost identical to our observed RV period of 1.5459 d, based on 26 observations. While a very close-in, tidally-locked planet could have the same orbital period as the stellar rotation period, given our other evidence, this seems highly unlikely in the case of Hubble I 4. The bisector analysis also suggests that we are looking at the effect of starspots.

The strongest evidence ruling out a companion is the wavelength dependence of the RV amplitude. This can be seen qualitatively in Figure 5, which shows the optical and NIR RVs together. A true reflex motion should have the same amplitude regardless of the wavelength of observation. The standard deviation of the optical RVs is 1068 \mps, while that of the NIR RVs is 383 \mps. Fitting a Keplerian orbit to the optical data yields a semi-amplitude of $1395\pm94$ \mps; similar analysis for the NIR data (with only the semi-amplitude as a free parameter) gives a semi-amplitude of $365\pm80$ \mps. We fit Keplerian orbits instead of sinusoids to allow the eccentricity to be a free parameter. We calculated the amplitude uncertainties with Monte Carlo trials with the individual RV errors scaled to produce $\chi^{2}=1$ for the Keplerian fits.

\begin{figure}
\plotone{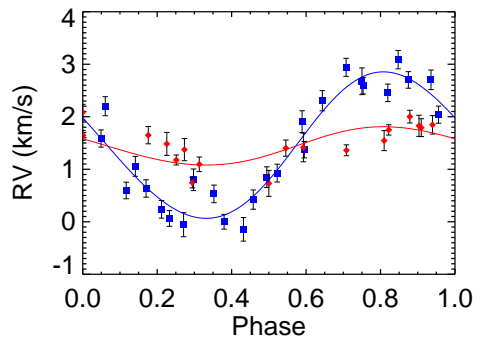}
\caption{Optical (\emph{blue squares}) and infrared (\emph{red diamonds}) phase-folded RVs for Hubble I 4, using a period of 1.5459 d. A Keplerian orbit was fit to the optical data (\emph{blue line}) yielding a semi-amplitude of $1395\pm94$ \mps. The same Keplerian orbit was also fit to the infrared RVs (\emph{red line}) with only the semi-amplitude ($365\pm80$ \mps) as a free parameter. A fixed offset was removed from the infrared RVs for ease of plotting.}  
\end{figure}

The spot-induced optical RV variability of Hubble I 4 is in line with previous observations of young stars. Saar \& Donahue (1997) explored the relationship between RV perturbations caused by starspots and convective inhomogeneities as these features are carried by rotation across the disk of a star and evolve over time. In general, they found these perturbations to increase with lower age, higher spot coverage, and higher $v$sin$i$. These relationships are consistent with the findings of Paulson et al. (2004) who searched for companions to stars in Hyades ($\sim$790 Myr) and Paulson \& Yelda (2006) who observed stars in $\beta$ Pictoris ($\sim$12 Myr), IC 2391 ($\sim$40 Myr), Castor ($\sim$200 Myr), and Ursa Majoris ($\sim$300 Myr). The authors measured RV variability of maximum amplitude $30-100$ \mps~for stars in Ursa Majoris and $250-600$ \mps~for stars in $\beta$ Pictoris. Stempels et al. (2007) observed periodic RV modulation of 2 \kmps~in the classical T Tauri star RU Lup, attributed to spots. In agreement with these observations, our younger Taurus-Auriga targets from our McDonald Observatory survey show much greater RV jitter than our older Pleiades targets (Mahmud 2011; Mahmud et al. 2011). This activity-age relationship can be seen clearly in Figure 6, which summarizes the above results.

\begin{figure}
\plotone{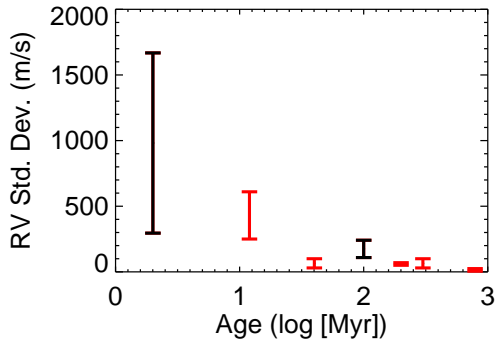}
\caption{Stars in younger clusters show greater optical RV jitter because of stronger magnetic fields. The \emph{bars} show the range of RV standard deviation in each cluster. From youngest to oldest, the clusters are Taurus-Auriga, $\beta$ Pictoris, IC 2391, Pleiades, Castor, Ursa Majoris, and Hyades. The \emph{red} data are from Paulson et al. (2004) and Paulson \& Yelda (2006) and have typical RV uncertainties of a few \mps~to $\sim$100 \mps; the \emph{black} data are from our McDonald Observatory survey (Mahmud 2011; Mahmud et al. 2011) and have typical RV uncertainties of $\sim$200 \mps.}
\end{figure}

Prato et al. (2008) presented similar results for the T Tauri stars DN Tau, V827 Tau, and V836 Tau. DN Tau was observed to have optical and NIR standard deviations of 440 \mps~and 144 \mps~respectively, V827 Tau had optical and NIR standard deviations of 1807 \mps~and 491 \mps~respectively, and V836 Tau had optical and NIR standard deviations of 742 \mps~and 149 \mps~respectively. While all three of these stars showed spot-induced RV modulation, the behavior of the NIR variability is not identical. V827 Tau and V836 Tau exhibited a reduction in velocity RMS from optical to NIR of a factor of $4-5$, while DN Tau and Hubble I 4 showed a reduction of only a factor of 3.

Both Hu{\'e}lamo et al. (2008) and Mart\'{i}n et al. (2006) performed similar multi-wavelength analysis for the young star TW Hya and the brown dwarf LP 944-20 respectively. They observed significant and dramatic decline in their H-band RV amplitudes compared to their optical ones. Unlike Hubble I 4 and V827 Tau, all of the examples discussed above (DN Tau, V836 Tau, TW Hya, LP 944-20) showed infrared RV variations similar in magnitude to the measurement uncertainties. In the case of Hubble I 4, however, the K-band RV variability is significantly greater than the measurement uncertainties. Figueira et al. (2010) reported a similar low-level infrared RV signal for TW Hya, with a reduction in amplitude from the optical of a factor of 3. These results demonstrate that with high RV precision (in the case of TW Hya) or high spot noise (in the case of Hubble I 4), it is possible to resolve the spot modulation even at longer wavelengths. Thus, any claims of companions around active stars based only on infrared RVs must be confirmed with observations at other wavelengths.

Reiners et al. (2010) investigated the effect of spots as a function of wavelength. Their spot simulations showed that the RV amplitude decreases as wavelength increases. However, this rate of decrease varies greatly depending on the temperature of the spot and photosphere. At low temperature contrasts ($\sim$200 K), this decrease can be a factor of 10 going from optical to NIR bands. However, the decrease can be significantly less for larger spot-photosphere temperature contrasts. Thus Hubble I 4 is likely to have a larger temperature contrast (i.e. cooler spots) compared to V827 Tau and V836 Tau.

As a first effort to explore whether a spot model can simultaneously produce the observed optical and infrared RV amplitudes of Hubble I 4, we modeled the RV variations with a single starspot using the technique described in Huerta et al. (2008). We used a stellar-disk integration model utilizing synthetic spectra. We derived our synthetic spectra using NextGen atmospheric models (Allard \& Hauschildt 1995). We divided the visible stellar disk into sections allowing for parts of it to be at a lower temperature than the rest of the photosphere, thus simulating the effect of a cool spot. We placed a spot of radius 25$^{\circ}$ at a colatitude of 45$^{\circ}$. We fixed $v$sin$i=14.6$ \kmps~and $i=8^{\circ}$; we derived these values using the stellar parameters of Hubble I 4 reported in Johns-Krull et al. (2004). In calculating our synthetic spectra, we assumed a photospheric temperature of 4200 K and a spot temperature of 3000 K. These temperature values are realistic for T Tauri stars; Joncour et al. (1994) detected a large polar spot 1600 K cooler than the surrounding photosphere in the T Tauri star HDE 283572. Our synthetic spectra yielded a spot/photosphere flux ratio of 0.022 in the optical (6275 \AA) and 0.597 in the infrared (22975 \AA). The resulting spot configuration at different rotational phases can be seen in Figure 7. Our disk integration produced simulated V-band and K-band spectra for this model at each rotational phase, using which we measured RVs in the same manner as with our actual Hubble I 4 optical spectra. These model RVs are shown in Figure 8.

\begin{figure}
\plotone{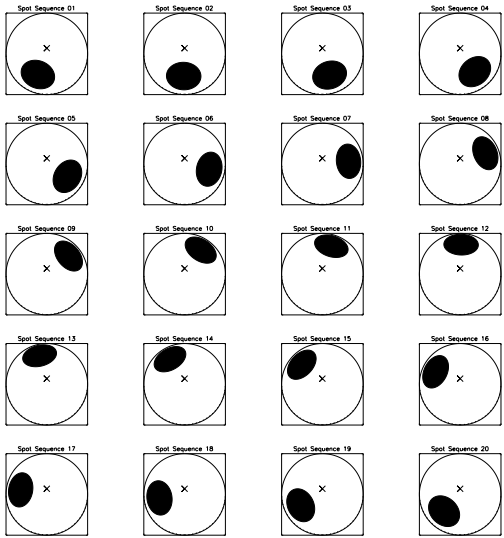}
\caption{Rotation sequence with a single spot of radius 25$^{\circ}$ at a colatitude of 45$^{\circ}$, with $i=8^{\circ}$. The \emph{cross} marks the pole.}
\end{figure}

\begin{figure}
%\plotone{f8.eps}
\includegraphics[scale = 0.5, trim = 0 60mm 0 0, clip]{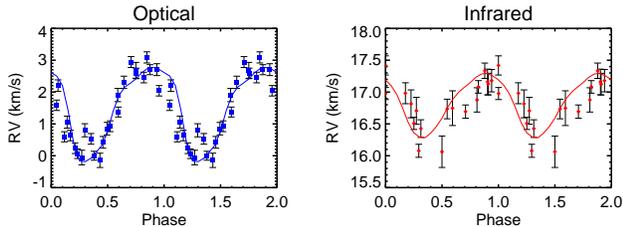}
\caption{Comparison of model RVs and measured RVs for Hubble I 4. The measured RVs are shown using \emph{blue squares} and \emph{red diamonds}. The model RVs are shown using \emph{lines}; they are from simulated spectra with a single circumpolar spot, with a photospheric temperature of 4200 K and a spot temperature of 3000 K.}
\end{figure}

Even with our simplistic model, the agreement with the measured RVs is good, especially in the optical. We are using our supporting optical photometry to do more sophisticated modeling of the spot configuration, temperature, and filling factor for Hubble I 4 and other T Tauri stars (to be reported in a forthcoming publication). Such results can be compared to measurements of magnetic field strengths (e.g. Johns-Krull 2007) and maps of magnetic field topology, photospheric brightness, and accretion-powered emission (e.g. Donati et al. 2010). 

Reiners et al. (2010) claimed that for spot-photosphere temperature contrasts of 1000 K or more, the NIR velocity amplitude is within a factor of 2 of that in the optical. While Reiners et al. (2010) did not consider the K-band, our results suggest that at some NIR wavelengths, spots with large temperature contrasts can still produce a change in velocity amplitude of close to a factor of 4, consistent with observations of stars such as Hubble I 4. Barnes et al. (2011) extended the work of Reiners et al. (2010) with more chaotic spot configurations. Such configurations are more likely to produce random RV jitter as opposed to a well-phased sinusoidal RV signal as seen in Hubble I 4. These stars will require more observations to recover a true companion-induced RV signal; however, they are less likely to produce a false detection. Additional multi-wavelength observations of both pre-main sequence T Tauri stars and main sequence M dwarfs are required to determine the exact relationship between mass, age, activity, and spots. This work will be crucial in determining the threshold of detection for planets around active stars. 

The optical RVs of Hubble I 4 show excellent phase coherence over 447 days. This suggests that the starspots are long-lived and that the magnetic field strength and topology on the stellar surface can stay more or less constant for many months. This longevity can provide clues about the origin of the magnetic fields. Magnetic fields on T Tauri stars might be the fossil remnants of the star formation process. This theory could help explain how two stars with very similar stellar parameters can nevertheless have different intrinsic fields leading to contrasting magnetospheric accretion rates (Johns-Krull et al. 1999). However, Donati et al. (2011) suggested that T Tauri magnetic fields are generated by dynamo processes rather than being fossil remnants. According to the authors, these dynamo fields are variable on timescales of a few years, thus explaining how stars with similar stellar parameters can have varying accretion rates. Multi-wavelength and long-baseline RV observations similar to the ones we have for Hubble I 4 might be useful in testing the cyclical nature of T Tauri magnetic fields.

\section{CONCLUSION}

Hubble I 4 shows significant and periodic RV variability resulting from the presence of cool surface spots. While many such RV measurements of active stars exist in the optical, this is one of the few cases in which the RV periodicity has been resolved in the infrared as well. There are two main implications: 

(1) RV planet searches around young or active stars (with a few large spots) cannot be conducted solely in the NIR. While the effect of spots is lessened at longer wavelengths, infrared RVs can still show periodicity similar to that expected for a companion. Observations at multiple wavelengths are essential to compare the RV amplitude.

(2) Active stars can show very different RV behavior as a function of wavelength depending on the physical properties of the spots. More observations of such active stars at multiple wavelengths, preferably simultaneously, are necessary to further investigate the structure of spots, and the relationship with magnetic fields and dynamos.

\acknowledgments

This work was partially supported by NASA Origins Grants 05-SSO05-86 and 07-SSO07-86; we also acknowledge the SIM Young Planets Key Project for research support. This research has made use of the SIMBAD database, operated at CDS, Strasbourg, France, and NASA's Astrophysics Data System. The authors wish to thank Wei Chen, Wilson Cauley, and Jennifer Blake-Mahmud for observing assistance at McDonald Observatory. We recognize the significant cultural role that Mauna Kea plays in the Hawaiian community and are grateful for the opportunity to observe there.\\

{\it Facilities:} \facility{IRTF (CSHELL)}, \facility{Smith (TS23)}.

\end{document}